# Wisdom of Crowds Detects COVID-19 Severity Ahead of Officially Available Data

Jeremy David Turiel [1], Delmiro Fernandez-Reyes [1], Tomaso Aste [1*]

*[1]* *Department of Computer Science, University College London, Gower Street, WC1E 6BT*

\* Corresponding author: Prof. Tomaso Aste; PhD. t.aste@ucl.ac.uk

## Abstract

During the unfolding of a crisis, it is crucial to determine its severity, yet access to reliable data is challenging. We investigate the relation between geolocated Tweet Intensity of initial COVID-19 related tweet at the beginning of the pandemic across Italian, Spanish and USA regions and mortality in the region a month later. We find significant proportionality between early social media reaction and the cumulative number of COVID-19 deaths almost a month later. Our findings suggest that 'the crowds' perceived the risk correctly. This is one of the few examples where the 'wisdom of crowds' can be quantified and applied in practice. This can be used to create real-time alert systems that could be of help for crisis-management and intervention, especially in developing countries. Such systems could contribute to inform fast-response policy making at early stages of a crisis.

**Introduction**

As of the 2nd of June 2020, a novel coronavirus pneumonia, COVID-19[1], has infected an estimated total of 6,194,000 individuals across virtually all the world's countries with 376,000 related deaths recorded globally[2].

Predicting the spread and forecasting the severity of the COVID- pandemic has become the focus of many research teams across the globe[3–6]. There is a shared agreement that forecasting the spread, growth and severity of the COVID-19 pandemic is a challenging task specially in our globally interconnected world. In this context, reliable prediction of contagion, growth and fatalities within countries and the regions in each country, before data is available and widely openly distributed, is essentially impossible. Despite this challenge, it is recognised that this knowledge is extremely useful in order to establish targeted confinement areas to contain virus spread more effectively while reducing the economic and social disruptions due to lockdown and social distancing strategies which in turn would also allow to allocate resources efficiently across regions.

Although forecasting a country's regional spread of COVID-19 severity and its associated mortality is critical to implement operational healthcare changes and epidemiological control measures, this is nearly impossible to achieve given the lack of readily available data at the beginning of the emerging SARS-CoV-2 pandemic. The collective knowledge of a crowd has been used successfully in similar challenging forecasting scenarios across social and data sciences where it has been established that collective opinions formed by a group of individuals can sometimes be more accurate than individual expert opinions. This has been called "the wisdom of crowds". As opposed to the common practice of using web-search engine queries that only indicate seeking knowledge patterns, we focused on assessing the wisdom of crowds as represented in the social media Twitter platform that was not available at the time of the past SARS outbreak. We therefore aggregated social media reaction, expressed by geolocated tweet intensity of COVID-19 related activity, from Italian, Spanish and United States geopolitical regions at the beginning of the pandemic as well as their regional mortality data.

To address the challenge of gaining these insights at the onset of a pandemic, when there is a lack of widely accessible regional-level official data, in the present work we analysed openly available data from twitter activity across different Italian, Spanish and United States regions to estimate crowd perception of the severity of the event.

In social sciences it has been established[7,8] that the collective aggregate opinion of a large number of non-experts can, in some contexts, outperform individual experts when the variable to be determined is not random and their individuals have some partial information and the ability to process it [7]. Social phycology studies have described crowd wisdom, opinion dynamics and collective knowledge by studying in which scenarios and ways crowds are wise [8]. Several studies have shown that social media wisdom of crowds, where user interactions are more frequent and opinion dynamics are heightened, are able to solve challenging forecasting tasks. Successful examples include improving Wikipedia articles [9]; predicting publicly traded securities stock prices[10]; the study of collective innovation in modern technological social networks [11] and; election results forecasting [12] among many others showing that crowd wisdom, opinion pooling and social media opinion dynamics constitute a powerful tool in forecasting tasks, even beyond experts' abilities. These methods work well especially when groups are large and connected opinion dynamics and communication allow crowds to process information[13]. Social media debate is the result of a complex process of information filtering where individuals gauge official information with local knowledge and confront their opinions openly. This process can degenerate in conspiracy theories and foster fake news, but it has been observed that on average the crowds can process information and weight reality in a rather accurate way[14]. While there are some examples of the use of the wisdom of crowds to estimate relevant variables that are otherwise hard to measure, there is no literature reporting the use of this collective knowledge gathered from social media tweets for assessing the spreading of mortality during an emerging pandemic. No previous study has shown that the collective wisdom of crowds during the initial attention COVID-19 social media peak, can strikingly predict the regional cumulative mortality a month ahead when there were not readily available mortality data sources.

Here we show that such "wisdom of crowds" has been able to predict, ahead of officially available data, COVID-19 infection severity across countries most affected by the pandemic. Our findings could underpin the creation of real-time novelty detection systems aimed at early reporting of SARS-like mortality and thus early activation of control measures in future pandemics. The wisdom of crowds could also be used to feed infection diseases explicit models with reliable data sourced locally from the exposed population when, at the early stages of a pandemic, there are no other sources of data available. The strength of the predictive association could be used to inform fast-response policy making where there is lack of readily available mortality data. As such it provides a scalable tool to increase preparedness and resilience in similar pandemic scenarios. Furthermore, the wisdom of crowds capability to infer the value of some variables otherwise unmeasurable can be used to refine explicit models.

**Results**

**Twitter activity as a proxy for crowd perception of the severity of COVID19.**

We have used readily accessible data from twitter activity across different Italian, Spanish and United States regions to estimate crowd perception of the severity of the event while it is unfolding in its early stages. We then related the intensity of social media reaction with the severity of the infection in the same region in terms of the cumulative number of deaths reported the following month. Our study focuses on Italy and Spain as these countries have been the most affected at the start of the pandemic followed more recently by the United States.

The number of active tweet users posting on COVID-19 per day[22,25], geolocated[24] and aggregated by the regions of Italy, Spain and the United States is shown in Figure 1a, 1b and 1c respectively. Country regions are coloured according to each country's geo-political areas. Figure 1 also shows the growth of positive SARS-CoV-2 cases nationwide as well as the cumulative number of deaths nationwide caused by COVID-19[15–17]. The cumulative mortality per geopolitical region of Italy, Spain and the United States is shown in Figure 2a, 2b and 2c respectively.

Using the Z-score method we identified for tweet intensity peaks the period 21st to 24th February 2020 for Italy, 24th to 26th February 2020 for Spain and 3rd to 4th March for the United States (Figure 1a, 1b, 1c). For the United States we observed a first peak in tweet intensity with Z > 3 around the 25th of February with no apparent endogenous cause such as change in confirmed cases or deaths (Figure 1c), it is then followed by a second peak that corresponds to an endogenous spike when the United States death toll starts to rise. This second United States tweet intensity peak (Figure 1c) was used for our analysis.

Note that Italian regional official data for the pandemic[15] was first available on the 24th February 2020 which is after the social media reaction (Figure 1a, Figure 2a). Moreover, at that time most Italian regions still reported no cases hindering the possibility of forecasting from official data (Figure 2a). The crowds therefore reacted on partial information (Figure 1a) that was not trivially obtainable from publicly available data. We observe an initial peak in late January (Figure 1a), perhaps due to the start of the epidemic in China, but with little differentiation between Italian regions. We then observe a second peak of interest from social media in late February (Figure 1a) that appears to be sparked by the endogenous growth of the infection in Italy being measured and reported. At the time (21st to 24th February 2020) only Italian nationwide epidemic data were available and regional or province breakdowns were only scattered across the news (Figure 1a, Figure 2a).

In order to show whether tweet intensity is related to the severity of COVID19 we plotted in log scale the cumulative number of deaths for each Italian (Figure 3a) and Spanish (Figure 3b) regions on the 7th of April 2020 and for the United States regions (Figure 3c) on the 14th of April 2020 against the mean tweet intensities at the beginning of the epidemic perception. We used the number of deaths, instead of population confirmed cases, as these are less dependent on the number of samples taken for SARS-CoV-2 testing in the wider population. Using population confirmed cases would have been highly dependent on the country testing strategies which would require a non-trivial rescaling. In other words, the dependence on testing strategies would have strengthened the relation with regional population spuriously. Figure 3 shows the proportionality between the mean tweet intensity at the beginning of the epidemic's perception, per region, and the number of deaths approximately one month forward. This figure demonstrates how the reaction on social media can correctly detect and rank the epidemic's impact on different regions one month ahead, when no official data was available in Italy and the data was insufficient for forecasting in other countries. This association is least noisy for Italy (Figure 3a) and Spain (Figure 3b) as these countries were severely affected very early on before the WHO declared the global pandemic. We note that the regions of Lombardy in Italy, Madrid in Spain and New York in the United States have the largest initial tweets intensity reactions and correspond to the most severely affected regions one month later. Note that the Lazio Italian region (Figure 3a) is an outlier due to politicians and central bodies tweeting from it as well as national geolocation defaulting to the capital. For Spain (Figure 3b), the region of "Castilla-La Mancha" was merged with Madrid as a large section of their population commute between the two and they are geographically nested. For the US the "District of Columbia", is over-represented with tweets from Washington D.C. (Figure 3c).

To assess the strength of the observed association shown in Figure 3 and to verify that the values of the epidemic are not trivially related to the size of the population in each region we compared three regression models: model-1, adjusted tweet intensity vs. log death cases; model-2, log population vs. log death cases and; model-3, adjusted tweet intensity and log population vs. log death cases.

We log-scaled the population to allow for a fair comparison as we notice a sub-linear relation to the number of deaths. Table 1 shows coefficients' significance as well as $R^2$ for the three weighted and non-weighted regression models for the three countries analysed in this study. From the data reported in Table 1 we observe that model-1 weighted regression is the most significant with the lowest p-values and the largest overall $R^2$ values for the Italy and Spain data. It has also a significant p-value and sizable $R^2$ for the United States. This indicates that tweet intensity is the most significant variable for the prediction of the number of deaths for Italy and Spain. For the United States we still observe that tweet intensity is a significant predictor for the number of death but results from unweighted regression with model-2 and model-3 reveal that the population of the regions is a better predictor.

We further assessed the strength and significance of the relation between the tweet intensity and the number of deaths by quantifying non-linear monotonic dependency with Spearman's Rho correlation (Table 2). We observe that the significance level for Spearman's Rho correlation for Italy is at 99% and both the United States and Spain are significant to 95% significance level (Table 2). This confirms that there is a significant relation between regional early tweet intensity s and the number of deaths in the respective regions for all three countries.

**Discussion**

The wisdom of crowds has been used successfully to estimate relevant variables that are otherwise hard to measure in several domains including the medical one[28]. Despite this, there is no literature reporting the use of the wisdom of crowds gathered from social media tweets for assessing the spreading of severity within an emerging pandemic where there is no readily access to mortality data.

Our study shows statistically significant evidence that COVID-19 related mean tweet intensity per region, at the first endogenous attention spike, is able to significantly forecast the spreading of COVID-19 severity, as measured by number of deaths, one month forward. In the case of Italy, the crowd's reaction with predictive power was recorded before any official regional contagion data was available. For Spain and the United States, the crowd still reacted when little data was available to make any forecast.

As the pandemic progressed, Italy, then rapidly followed by Spain, were the first countries affected with extremely high COVID-19 associated mortality. Italy and Spain were therefore less influenced by discussions about the general global status of the pandemic, as less attention was present at the time. This allowed us to analyse the reaction from social media crowds with less biases. Moreover, Italy is made up of a good number of regions of comparable size with good social media usage as well as good official data for social media usage.

We show that the intensity of COVID-19 related twitter activity is able to correctly identify the localities most affected by the pandemic in each of the considered countries. These localities are Lombardy, Madrid and New York for Italy, Spain and the United States respectively (Figure 3). These geo-political and administrative localities have a striking social media reaction (Figure 1, Figure 3). This suggests that the initial reaction of users on social media had efficiently processed data scattered throughout news channels, merged it with local information and performed an accurate risk assessment which is observable in the social media intensity reaction. We highlight in particular for Italy that Emilia Romagna and Veneto did not seem to be less affected than Lombardy at the beginning of the epidemic spread, however, nonetheless, crowd wisdom seems to have combined different information sources to highlight the perception of a greater danger in Lombardy (Figure 2a, Figure 3a). In Italian and Spanish regions, the crowds demonstrated a remarkable ability to predict the severity of COVID-19 impact at regional levels before the availability of official data. The regression results demonstrate that the intensity of COVID-19 related tweeds is a better predictor than the population size obtaining goodness of fit $R^2$ values that are almost twice. For the United States we also demonstrated a significant predictability power of the tweet intensity, however in this case the regional population size is a better regressor. We might note that spreading of the pandemic in the United States started later when an amount of exogenous information was already circulating in the social media, furthermore the United States have a much wider variety of climate, culture, political guidance, population density, and total population throughout states that leads to a more difficult detection of the phenomenon (Figure 3c).

The crowd's reaction to COVID-19 spreading measured through tweet intensity on regional basis is a complex quantity rich of information. People react to both official information and to local knowledge gathered at personal level. Tweets are a process involving both sharing and comparing such information which includes a level of collective processing and assessing of the reliability of the source. It has been already recognized in the literature that such a process can be in average very accurate. It should not be therefore surprising that the crowd interest measured through COVID-19 related tweet intensity can result in an appropriate estimate of the local severity of the epidemics.

The practical relevance of our results consists in the demonstration that tweet intensity can be used for forecasting. At the beginning of an epidemics it is extremely difficult to have precise information and therefore modelers and public officials are obliged to use general statistical quantities to produce forecasts and consequently implement decisions. Our work shows that the information locally available to the population permeates through twitter and social media and can be made available to modellers and policymakers at the early stages of a crisis when it is most needed.

**Materials and Methods**

### COVID-19 and population data sources.

We obtained COVID-19 spreading and casualties time series aggregated by region from the official department of health website or repository for Italy[15] and Spain[16]. For the United States we used the readily available New York Times dataset[17] which contains regional level data as well as an interactive package for live monitoring.

For Italy we obtained both population data per region[18] and social media usage statistics per region[19], both updated as of 2019, from Istituto Nazionale di Statistica (ISTAT). We obtain regional population data for Spain from Istituto Nacional de Estadística (INE) as of 2019[20] and for the United States from the United States Census Bureau[21].

**Social media data crawling and processing.**

We obtained all twitter data from an early COVID-19 twitter data repository[22] available before Twitter started providing access to its own COVID-19 dataset. Tweets were collected from the 21st of January 2020 using Twitter's Application Programming Interface (API)[23]. The Twitter repository is updated as new meaningful COVID-19 related words emerge [23]. Using tweet unique identifiers (IDs), we retrieved all their corresponding information (text, date, user and user data) from the repository via the "*twarc*" package (https://github.com/DocNow/twarc). This process is commonly referred to as tweet hydration. To measure the number of unique active users per day in each of the regions of Italy, Spain and the United States, we geolocated tweets with the HERE Geocoder service[24] and obtained country and region categorization for over 50% of the tweets. In order to facilitate dataset reading and tabulation we used Google's OpenRefine API[25] that efficiently groups hourly tweet data by day and convert nested dictionary type files in JavaScript Object Notation (JSON) to simple tabular comma separated values.

We defined tweet volume as the number of unique users active in a region per day, discussing COVID-19 related topics. This criterion is applied, rather than merely counting tweets, in order to measure the population's attention and news spreading in the crowd, while correcting for overrepresented twitter activity from certain users. To account for different sizes of regions' populations we calculate tweet intensity by normalizing the tweet volume by region population in the case of Spain and the United States or by social media active population in the case of Italy. Indeed, Italy was the only country for which we were able to obtain specific social media usage data per region.

**Identification of tweet intensity peaks**

In order to identify the beginning of the social media reaction to the epidemic in each country we computed the z-score on the tweet activity time series on two weeks rolling window. We used Z>3 as the identifier of the social media reaction peak.

**Regression analyses**

Weighted and non-weighted linear regressions models were carried out using the python package "*statsmodels.api*"[26] (python version 3.7). The Weighted Least Squares (WLS) function was uses for weighted linear regression (with intercept) and the Ordinary Least Squares (OLS) function was used for unweighted linear regression (with intercept). The weighted regressions account for variability in the data by weighting the error proportionally to the square root of region population. This adjustment accounts for error measurements both in number of deaths and twitter volume. Regression *p*-values are calculated by the "*.summary()*" function of the "*.fit()*" method for the OLS and WLS functions. This derives from Chi Squared survival function of "*scipy.stats*"[27] (scipy.stats.chi2.sf()) applied to the test statistics, which is assumed to be Chi Square distributed.

**Correlation analyses**
Spearman's rank correlation coefficient was calculated using the *"spearmanr"* function of *"scipy.stats"*[27]. In order to assess the strength of this correlation we compare it with null-hypothesis correlation values for random shuffling of the data and obtain quantile confidence intervals. We hence obtain confidence levels for comparison with the random null model.

## Acknowledgments

**General**: The authors acknowledge Fabio Caccioli, Licia Capra, Giacomo Livan and Simone Righi for useful feedback and discussions.

**Funding:** TA and JT acknowledge support from the EC Horizon 2020 FIN-Tech and EPSRC (EP/L015129/1). TA acknowledges support from ESRC (ES/K002309/1); EPSRC (EP/P031730/1) and; EC (H2020-ICT-2018-2 825215). DFR acknowledges support from EPSRC (EP/P028608/1). The funders had no role in the writing of the manuscript nor in the decision to submit it for publication. We have not been paid to write this article by a pharmaceutical company or any other agency. All authors including the corresponding senior authors have had full access to all the data and had the final responsibility for the decision to submit for publication.

**Author contributions:** JT, DFR, TA conceived the study. JT carried out data crawling, processing and implemented the analyses. JT, DFR, TA carried out analysis of results and wrote the manuscript. All authors have approved the manuscript.

**Competing interests:** The authors declare no competing interests.

**Data and materials availability:** During manuscript assessment and review the data can be accessed at: https://figshare.com/s/b344e72688e047d2d65d
Data openly available upon publication in Turiel, Jeremy; Fernandez-Reyes, Delmiro; Aste, Tomaso (2020): The Wisdom of the Crowd in Assessing the Spread of COVID-19 Associated Mortality. University College London.
Dataset. https://doi.org/10.5522/04/12317339.

## References and Notes

1. Li Q, Guan X, Wu P, *et al.* Early transmission dynamics in Wuhan, China, of novel coronavirus-infected pneumonia. N. Engl. J. Med. 2020. DOI:10.1056/NEJMoa2001316.
2. WHO Health Emergencies Program. WHO Coronavirus Disease (COVID-19) Dashboard. World Heal. Organ. 2020. https://covid19.who.int (accessed May 14, 2020).
3. Zhang J, Litvinova M, Wang W, *et al.* Evolving epidemiology and transmission dynamics of coronavirus disease 2019 outside Hubei province, China: a descriptive and modelling study. *Lancet Infect Dis* 2020.
4. Chinazzi M, Davis JT, Gioannini C, *et al.* Preliminary assessment of the International Spreading Risk Associated with the 2019 novel Coronavirus (2019-nCoV) outbreak in Wuhan City. *Lab Model Biol Soc--Techn Syst* 2020.
5. Roosa K, Lee Y, Luo R, *et al.* Real-time forecasts of the COVID-19 epidemic in China from February 5th to February 24th, 2020. *Infect Dis Model* 2020; **5**: 256–63.

6   Grasselli G, Pesenti A, Cecconi M. Critical care utilization for the COVID-19 outbreak in Lombardy, Italy: early experience and forecast during an emergency response. *JAMA* 2020.

7   Wagner C, Vinaimont T. Evaluating the wisdom of crowds. *Proc Issues Inf Syst* 2010.

8   Mannes AE, Larrick RP, Soll JB. The social psychology of the wisdom of crowds. 2012.

9   Golub B, Jackson MO. Naive learning in social networks and the wisdom of crowds. *Am Econ J Microeconomics* 2010; **2**: 112–49.

10  Azar PD, Andrew W Lo. The wisdom of Twitter crowds: Predicting stock market reactions to FOMC meetings via Twitter feeds.". *J Portf Manag* 2016; **42**: 123–34.

11  Kozinets R V, Hemetsberger A, Schau HJ. The wisdom of consumer crowds: Collective innovation in the age of networked marketing. *J macromarketing* 2008; **28**: 339–54.

12  Olsson H, de Bruin WB, Galesic M, Prelec D. Harvesting the wisdom of crowds for election predictions using the Bayesian Truth Serum. 2019.

13  Bassamboo A, Cui R, Moreno A. Wisdom of Crowds in Operations: Forecasting Using Prediction Markets. *Available SSRN 2679663* 2015.

14  Pennycook G, Rand DG. Fighting misinformation on social media using crowdsourced judgments of news source quality. *Proc Natl Acad Sci U S A* 2019. DOI:10.1073/pnas.1806781116.

15  del Consiglio dei Ministri - Dipartimento della Protezione Civile P. Dati COVID-19 Italia. GitHub Repos. 2020.

16  de Sanidad Im. Situación de COVID-19 en España. 2020.

17  The New York Times. Data from The New York Times, based on reports from state and local health agencies. 2020.

18  ISTAT IN di S. Popolazione residente al 1° gennaio. 2020.

19  ISTAT IN di S. Internet: accesso e tipo di utilizzo: Attività svolte su internet - reg. e tipo di comune. 2020.

20  de Estadística IN. Cifras oficiales de población resultantes de la revisión del Padrón municipal a 1 de enero - Resumen por comunidades autónomas. 2019.

21  U.S. Census Bureau PD. Table 1. Annual Estimates of the Resident Population for the United States, Regions, States, and Puerto Rico: April 1, 2010 to July 1, 2019 (NST-EST2019-01). 2019.

22  Chen E, Lerman K, Ferrara E. COVID-19: The First Public Coronavirus Twitter Dataset. *arXiv Prepr arXiv200307372* 2020.

23  Twitter I. Twitter Developer API. https://developer.twitter.com/en/docs (accessed May 8, 2020).

24  HERE Developer. HERE Geocoder API. https://developer.here.com/documentation/geocoder/dev_guide/topics/what-is.html (accessed April 15, 2020).

25  Huynh D. OpenRefine. https://github.com/OpenRefine/OpenRefine.

26  Seabold S, Perktold J. statsmodels: Econometric and statistical modeling with python. In: 9th Python in Science Conference. 2010.

27  Virtanen P, Gommers R, Oliphant TE, *et al.* SciPy 1.0: Fundamental Algorithms for Scientific Computing in Python. *Nat Methods* 2020; **17**: 261–72.

28   Beyrer C, Baral SD, Van Griensven F, *et al.* Global epidemiology of HIV infection in men who have sex with men. Lancet. 2012. DOI:10.1016/S0140-6736(12)60821-6.

**Figures and Tables**

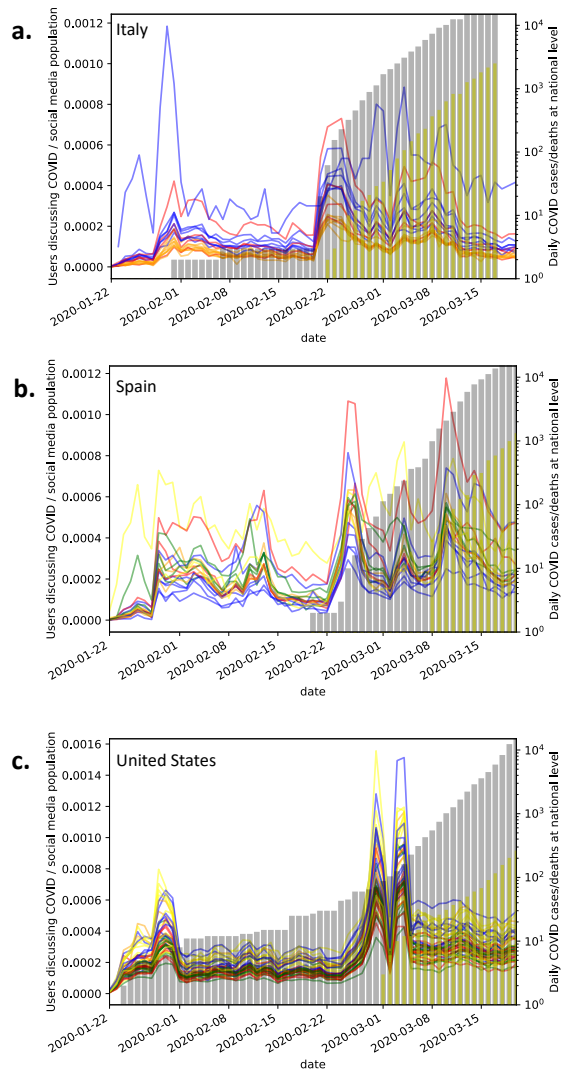

**Figure 1. Plots of COVID-19 related twitter activity superimposed to number of confirmed COVID-19 cases and cumulative number of deaths nationwide.** Twitters are geolocated and aggregated by geopolitical regions for **a.** Italy; **b.** Spain and **c.** United States. For each country we group and color code regions by geolocation in the following way: **a.** Italy: Northern regions (blue), Central regions (red) and Southern regions (orange) **b.** Spain: Northeastern regions (blue), Northwestern regions (green), Central regions (red), Southern regions (orange), Autonomous cities (yellow) **c.** United States: Northeastern states (blue), Southeastern states (green), Midwestern states (red), Southwestern states (orange), Western states (yellow). The growth of confirmed COVID-19 cases nationwide (cumulative) is represented by grey bars while the cumulative number of COVID -19 deaths nationwide is represented by yellow bars.

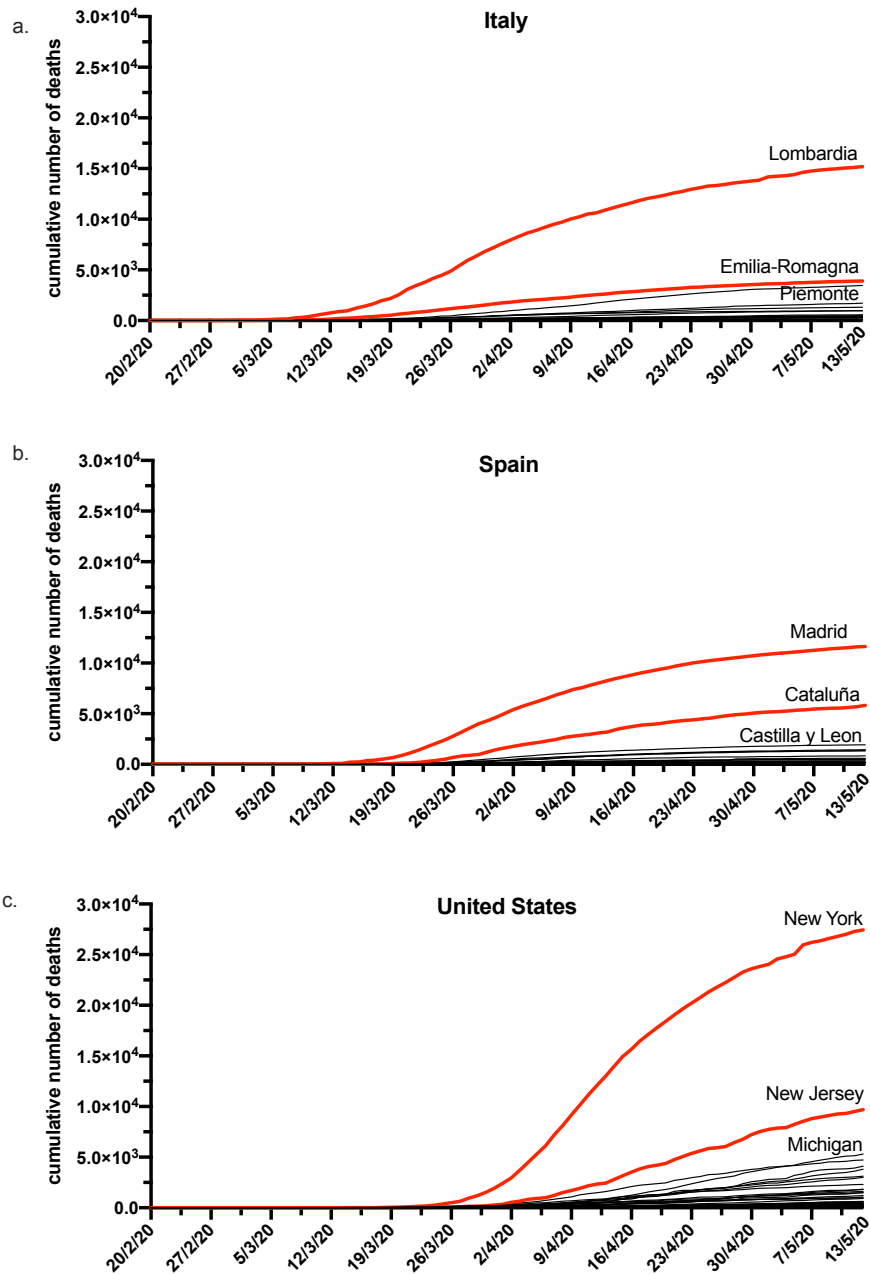

**Figure 2. Plots of the cumulative number of deaths per geopolitical region** for: a. Italy; b. Spain; c. United States

**Figure 3. Demonstration that the reaction on social media can correctly detect and rank the epidemic's impact on different regions one month ahead.** The y-axis reports the cumulative number of deaths one month forward and the x-axis reports mean tweet intensity at the initial attention peak per geopolitical region of: a. Italy; b. Spain and c. United States. The diameters of the circles are proportional to the population of the region.

| Regression Model | Country | Weighted | | | Non-Weighted | | |
|---|---|---|---|---|---|---|---|
| | | $p$ Adjusted Tweets | $p$ log (Population) | $R^2$ | $p$ Adjusted Tweets | $p$ log (Population) | $R^2$ |
| **Model-1** Adjusted Tweets vs. log (Death Cases) | Italy | $2 \times 10^{-7}$ | - | 0.798 | $5 \times 10^{-4}$ | - | 0.489 |
| | Spain | $1 \times 10^{-5}$ | - | 0.684 | $4 \times 10^{-2}$ | - | 0.402 |
| | US | $5 \times 10^{-6}$ | - | 0.340 | $3 \times 10^{-2}$ | - | 0.073 |
| **Model-2** log (Population) vs. log (Death Cases) | Italy | - | $9 \times 10^{-4}$ | 0.456 | - | $9 \times 10^{-3}$ | 0.431 |
| | Spain | - | $8 \times 10^{-4}$ | 0.482 | - | $2 \times 10^{-3}$ | 0.475 |
| | US | - | $9 \times 10^{-4}$ | 0.189 | - | $1 \times 10^{-13}$ | 0.680 |
| **Model-3** Adjusted Tweets and log (Population) vs. log (Death Cases) | Italy | $3 \times 10^{-6}$ | $1 \times 10^{-2}$ | 0.857 | $3 \times 10^{-4}$ | $8 \times 10^{-4}$ | 0.737 |
| | Spain | $7 \times 10^{-4}$ | $4 \times 10^{-2}$ | 0.748 | $3 \times 10^{-1}$ | $1 \times 10^{-1}$ | 0.477 |
| | US | $1 \times 10^{-3}$ | $3 \times 10^{-1}$ | 0.343 | $3 \times 10^{-1}$ | $1 \times 10^{-12}$ | 0.681 |

**Table 1. Per country coefficient significance and R2 values for weighted and unweighted regression models.** Regression models explored: tweets intensity vs. log death cases; log population vs. log death cases and; tweets intensity and log population vs. log death cases.

| | Country | Empirical correlation value | Null hypothesis correlation quantiles | | |
|---|---|---|---|---|---|
| | | | *90%* | *95%* | *99%* |
| **Adjusted Tweets vs. log (Death Cases)** | Italy | 0.66 | 0.31 | 0.39 | 0.53 |
| | Spain | 0.51 | 0.34 | 0.42 | 0.58 |
| | US | 0.31 | 0.18 | 0.24 | 0.33 |

**Table 2.** Spearman's rank correlation coefficient values for empirical values of each country and corresponding null-models significance levels.